\documentclass[twocolumn,aps,a4paper,showpacs,amsfonts,amsmath,amssymb]{revtex4}
\usepackage{makeidx}
\usepackage{graphicx}
\usepackage{epsfig}

\begin{document}

\title{Non-stationary Feller process with time-varying coefficients} 

\author{Jaume Masoliver}
\email{jaume.masoliver@ub.edu}
\affiliation{Departament de F\'{\i}sica Fonamental, Universitat de Barcelona,\\
Diagonal, 647, E-08028 Barcelona, Spain}

\date{\today}

\begin{abstract}

We study the non-stationary Feller process with time varying coefficients. We obtain the exact probability distribution exemplified by its characteristic function and cumulants. In some particular cases we exactly invert the distribution and 
achieve the probability density function. We show that for sufficiently long times this density approaches a Gamma distribution with time-varying shape and scale parameters. Not far from the origin the process obeys a power law with an exponent dependent of time, thereby concluding that accessibility to the origin is not static but dynamic. We finally discuss some possible applications of the process.

\end{abstract}
\maketitle

\section{Introduction}
\label{intro}

The Feller process is a one-dimensional diffusion  with linear drift and linear diffusion coefficient vanishing at the origin.  One of its most distinctive characteristics is that it never attains negative values. This, along with the fact that the process amounts to a continuous representation of a number of branching and birth-dead processes, have made Feller process an ideal candidate for modeling many phenomena in physical and social sciences \cite{maso12,gan}. Examples span  from theoretical biology and neurobiology \cite{ricciardi77,gerstner,lanska,lansky,mckane}, population growth \cite{capocelli_ricciardi,capocelli,murray,azaelenature,azaele}, radiation physics \cite{radiation} and even economics and financial markets  \cite{cox,hull,heston,yakovenko,maso08,maso09}. From a more formal point of view, the process is also valuable because some one-dimensional diffusions can be mathematically transformed into the Feller process, as it was proved by Capocelli and Ricciardi years ago \cite{capocelli76}. 

In its original formulation \cite{feller1}  the parameters are time independent and the process is stationary. This means that it is  invariant under time translations and that, as time progresses, the probability distribution tends towards a stationary value given by the Gamma distribution \cite{maso12}. However, as countless situations show, stationary processes are a convenient idealization and many empirical studies indicate that real data susceptible to be modeled by diffusion processes are not fully stationary \cite{doyne}. 

There are, however, very few attempts at addressing this question for the Feller process (as well as for other diffusions). One exception is, to our knowledge, the recent work by Gan and Waxman \cite{gan} where the authors, using the method of spectral decomposition, obtain an explicit solution for the distribution function of the process with time varying coefficients. Although they achieve the solution only in the special and singular case when the linear drift also vanishes at the origin. This is rather restricting because, as we will see below, the value of the drift at the origin determines where the process tends to in the course of time. Fixing this value to zero limits the scope and applications of the solution (even though it is valuable in population dynamics when one is faced with possible extinctions \cite{capocelli,azaele}). 

In this paper we address the study of the Feller process in the most possible general form. The main result is knowing the exact probability distribution materialized by the Laplace transform of the probability density function, that is, by its characteristic function. The transformed density is inverted in some particular but relevant cases (including that of the original process). In addition we also get some asymptotic expressions valid for large to moderate times showing the emergence of the Gamma distribution but with time-varying shape and scale parameters. 

Another significant aspect is the behavior of the process near the origin which, in turn, determines whether or not the origin is accessible. In the stationary case accessibility depends on the value taken by a constant ratio between two parameters governing the deterministic and random components of the dynamics of the process \cite{maso12,feller1}. We will see that for the non-stationary process the behavior of the probability density function near the origin is given by a power law with a time-dependent exponent which implies that the likelihood of attaining the origin is not fixed but changing with time. 

The paper is organized as follows. In Sec. \ref{dynamics} we study the dynamics and the general properties 
of the non-stationary process. In Sec. \ref{cf} we obtain the exact probability distribution through the knowledge of its characteristic function and cumulants. Section \ref{special} is devoted to obtaining the exact form of the probability density function for some special cases.  In Sec. \ref{assym} we get approximate expressions valid for moderate to long times. Section \ref{origin} focusses on the behavior of the probability density near the origin and its attainability by the process. A short summary of the main results with few possible applications is in Sec. \ref{conclusions} and more technical details are in several  Appendices.

\section{Dynamics and general properties}
\label{dynamics}

In its most general form the Feller process can be described by the following Langevin equation:
\begin{equation}
\frac{dX(t)}{dt}=-[\alpha(t)X(t)-\beta(t)]+k(t)\sqrt{X(t)}\xi(t),
\label{langevin_1}
\end{equation}
where $\xi(t)$ is the derivative of the standard Wiener process. That is, $\xi(t)$ zero-mean Gaussian white noise 
$$
\langle\xi(t)\rangle=0, \qquad \langle\xi(t)\xi(t')\rangle=\delta(t-t'). 
$$
All stochastic differential equations of this paper are interpreted in the sense of It\^o. 

Although from a mathematical point of view the parameters of the process are given by arbitrary functions of time, from a physical point of view it is more convenient to assume that $\alpha(t)>0$, $\beta(t)\geq 0$ and $k(t)>0$ are positive and smooth functions. This will be one of our basic assumptions for the rest of the paper.  

Note that $\alpha(t)$ has dimensions of (time)$^{-1}$ and sets a dimensionless time scale defined by
\begin{equation}
\tau=\int_{t_0}^t \alpha(t')dt',
\label{tau}
\end{equation}
where $t_0$ is any initial time. Observe that $t=t_0$ corresponds to $\tau=0$ and also that $\tau=\tau(t)$ is an increasing function of the original time scale because $d\tau/dt=\alpha(t)>0$ $(t\geq t_0)$. In other words, ``future maps into future" (and past into past). 

For the rest of the paper we also assume that $\alpha(t)$ is such that 
$$
\lim_{t\to\infty}\int_{t_0}^t \alpha(t') dt'=\infty.
$$
That is to say, $t\to\infty$ implies $\tau\to\infty$ and vice versa. This is our second and last restriction on the parameters of the process. 

In terms of this new time scale we write the stochastic equation (\ref{langevin_1}) as
$$
\frac{dX(\tau)}{d\tau}=-\left[X(\tau)-\frac{\beta(t(\tau))}{\alpha(t(\tau))}\right]+
\frac{k(t(\tau))}{\alpha(t(\tau))}\sqrt{X(\tau)}\xi(t(\tau)),
$$
where $t=t(\tau)$ is the inverse function implicitly defined in Eq. (\ref{tau}). Defining the rescaled parameters
\begin{equation}
m(\tau)\equiv \frac{\beta(t(\tau))}{\alpha(t(\tau))}, \qquad D(\tau)\equiv \frac{k^2(t(\tau))}{2\alpha(t(\tau))},
\label{rescaled_par}
\end{equation}
and the rescaled noise  
\begin{equation}
\eta(\tau)\equiv\frac{1}{\sqrt{\alpha(t(\tau))}}\xi(t(\tau)),
\label{rescaled_noise}
\end{equation}
the Langevin equation for $X(\tau)$ reads 
\begin{equation}
\frac{dX(\tau)}{d\tau}=-\bigl[X(\tau)-m(\tau)\bigr]+\sqrt{2D(\tau)X(\tau)} \eta(\tau).
\label{langevin_2}
\end{equation}

The rescaled noise $\eta(\tau)$ is zero-mean Gaussian white noise:
\begin{equation}
\langle\eta(\tau)\rangle=0, \qquad \langle\eta(\tau_1)\eta(\tau_2)\rangle=\delta(\tau_1-\tau_2). 
\label{correlation}
\end{equation}
Indeed, that $\eta(\tau)$ is Gaussian noise with zero mean is evident from Eq. (\ref{rescaled_noise}). We next show that $\eta(\tau)$ is delta correlated. From definition (\ref{rescaled_noise}), we have
$$
\langle \eta(\tau_1)\eta(\tau_2)\rangle=\frac{1}{\sqrt{\alpha(\tau(t_1))\alpha(\tau(t_2))}}\delta[t(\tau_1)-t(\tau_2)].
$$
Let us recall the standard property of the delta function, $\delta[g(x)]=\delta(x-x_0)/|g'(x_0)|$, where $g(x)$ is any differentiable function and $x_0$ is the solution to $g(x_0)=0$. Then, since $t(\tau_1)=t(\tau_2)$ is equivalent to 
$\tau_1=\tau_2$ [we assume that $t(\tau)$ is an univariate function], we write
\begin{eqnarray*}
\delta[t(\tau_1)-t(\tau_2)]&=&\frac{1}{|t'(\tau_2)|}\delta(\tau_1-\tau_2)=|\tau(t_2)|\delta(\tau_1-\tau_2)\\
&=&
\alpha(\tau(t_2))\delta(\tau_1-\tau_2),
\end{eqnarray*}
where the prime refers to the derivative and $t(\tau)$ and $\tau(t)$ are inverse functions for which 
$t'=1/\tau=1/\alpha$ [cf. Eq. (\ref{tau})]. Hence
$$
\langle \eta(\tau_1)\eta(\tau_2)\rangle=\frac{\alpha(t(\tau_2))}{\sqrt{\alpha(\tau(t_1))\alpha(\tau(t_2))}}\delta(\tau_1-\tau_2)=
\delta(\tau_1-\tau_2),
$$
which proves Eq. (\ref{correlation}). 

In the representation given by Eq. (\ref{langevin_2}) the process has a state and time dependent diffusion 
$D(X,\tau)=2D(\tau)X$, which for large values of $X$ enhances the effects of noise (with an intensity depending of time) while as $X$ approaches zero the effect of noise diminishes. Hence, when the process reaches the origin, the drift drags it toward the asymptotic value: 
\begin{equation}
m_\infty=\lim_{\tau\to\infty}m(\tau).
\label{m_infty}
\end{equation}

Indeed, near the origin the process (\ref{langevin_2}) approximates to the deterministic process $dx/d\tau=-[x-m(\tau)]$ and from its solution we readily see that $x(\tau)\to m_\infty$ as $\tau\to\infty$. Consequently, if $m(\tau)\geq 0$ the process, starting at some positive value, cannot reach the negative region and the Feller process is nonnegative [otherwise the noise term in Eq. (\ref{langevin_2}) would become imaginary]. 

Moreover, the asymptotic value $m_\infty$ coincides with the stationary average of the process,
\begin{equation}
m_\infty=\lim_{\tau\to\infty}\langle X(\tau)\rangle.
\label{average}
\end{equation}
In effect, the formal solution of Eq. (\ref{langevin_2}) such that $X(0)=x_0$ is
\begin{eqnarray*}
X(\tau)=e^{-\tau}\biggl[x_0&+&\int_0^\tau e^{\tau'}m(\tau')d\tau'\\
&+& \int_0^\tau e^{\tau'}\sqrt{2D(\tau')X(\tau')}\eta(\tau')d\tau'\biggr].
\end{eqnarray*}
As is well known, in the It\^o interpretation the output process $X(\tau)$ and the input white noise $\eta(\tau)$ are uncorrelated. Hence
$$
\left\langle\sqrt{X(\tau)}\eta(\tau)\right\rangle=\left\langle\sqrt{X(\tau)}\right\rangle \langle\eta(\tau)\rangle=0,
$$
whence
$$
\langle X(\tau)\rangle=e^{-\tau}\left[x_0+\int_0^\tau e^{\tau'}m(\tau')d\tau'\right],
$$
which can be written as
$$
\langle X(\tau)\rangle=x_0e^{-\tau}+\int_0^\tau e^{-s}m(\tau-s)ds,
$$
and the limit $\tau\to\infty$ proves Eq. (\ref{average}). Let us observe that when $m(\tau)=m$ is constant, $m_\infty=m$, and the process tends to $m$. This is the reason why in many setting (such as, for example, economics  \cite{cox,hull,heston,yakovenko,maso08,maso09}) $m$ is called the ``normal level'' of the process.

We, therefore, see that for the non-stationary Feller process, and under rather general circumstances ({\it i.e.,} $m(\tau)\geq 0$), the origin is a singular boundary that the process cannot cross. A closely related problem is whether or not the origin is attainable, in other words, whether the value $X=0$ can or cannot be reached. This is a key issue in many practical situations (for instance, in population dynamics where attaining the origin signifies annihilation). We will prove later on that the answer to this question is time dependent and determined by a varying parameter balancing deterministic motion and fluctuations.  The problem of classifying the different types of boundaries appearing in diffusion processes was thoroughly studied by Feller himself in the early 1950s and we refer the reader to the literature for a more complete account on this topic \cite{feller2,feller3,gardiner}.

\section{Probability distribution and cumulants}
\label{cf}

Our main objective  is knowing the probability distribution of the non-stationary Feller process (\ref{langevin_1}). In this section we will determine the distribution of probability through  the characteristic function (CF) as well as  the cumulants. 

In the time scale defined by Eq. (\ref{tau}), the probability density function (PDF) of the process $X(\tau)$,
$$
p(x,\tau|x_0)dx={\rm Prob}\bigl\{x<X(\tau)<x+dx|X(0)=x_0\bigr\},
$$
obeys the Fokker-Planck equation (FPE) \cite{gardiner}
\begin{equation}
\frac{\partial p}{\partial\tau}=\frac{\partial }{\partial x}\Bigl(\bigl[x-m(\tau)\bigr]p\Bigr)+D(\tau) \frac{\partial^2}{\partial x^2}(xp),
\label{fpe}
\end{equation}
with initial condition
\begin{equation}
p(x,0|x_0)=\delta(x-x_0).
\label{initial}
\end{equation}

Recall that $x=0$ is a singular boundary that the process cannot cross. A sufficient condition for this to happen is that the probability flux through $x=0$ is zero. We thus look for solutions of the initial-value problem (\ref{fpe}) and (\ref{initial}) that meet this condition:
\begin{equation}
\lim_{x\to 0}J(x,\tau|x_0)=0,
\label{flux}
\end{equation}
where
\begin{equation}
J(x,\tau|x_0)=\left[x-m(\tau)+D(\tau)\frac{\partial }{\partial x} x\right]p(x,\tau|x_0),
\label{flux_def}
\end{equation}
is the flux of probability through $x$ at time $\tau$. Note that in terms of the flux the FPE (\ref{fpe}) can be written as
\begin{equation}
\frac{\partial p}{\partial\tau}=\frac{\partial }{\partial x}J(x,\tau|x_0).
\label{fpe_flux}
\end{equation}

We will now proceed to obtain the exact probability distribution via the characteristic function of the process, the latter defined as the Laplace transform of the PDF with respect to $x$ (bear in mind that $x\geq 0$):
\begin{equation}
\hat p(\sigma,\tau|x_0)=\int_0^\infty e^{-\sigma x} p(x,\tau|x_0) dx.
\label{LT}
\end{equation}
The Laplace transform of Eq. (\ref{fpe_flux}) under condition (\ref{flux}) reads
$$
\frac{\partial\hat p}{\partial\tau}=\sigma\int_0^\infty e^{-\sigma x} J(x,\tau|x_0),
$$
and substituting for Eq. (\ref{flux_def}) we have
\begin{eqnarray*}
\frac{\partial\hat p}{\partial\tau}&=&\sigma\Biggl\{\int_0^\infty xe^{-\sigma x} p(x,\tau|x_0)dx- m(\tau)\hat p(\sigma,\tau|x_0)\\
&+&
D(\tau)\int_0^\infty e^{-\sigma x} \frac{\partial }{\partial x}[xp(x,\tau|x_0)]dx\Biggr\},
\end{eqnarray*}
but
$$
\int_0^\infty xe^{-\sigma x} p(x,\tau|x_0)dx=-\frac{\partial}{\partial\sigma}\hat p(\sigma,\tau|x_0),
$$
and 
$$
\int_0^\infty e^{-\sigma x} \frac{\partial }{\partial x}[xp(x,\tau|x_0)]dx=
-\sigma\frac{\partial}{\partial\sigma}\hat p(\sigma,\tau|x_0).
$$
Hence
$$
\frac{\partial\hat p}{\partial\tau}=-\sigma\left[\frac{\partial\hat p}{\partial\sigma}+
m(\tau)\hat p+\sigma D(\tau)\frac{\partial\hat p}{\partial\sigma}\right].
$$

We, therefore, obtain the following partial differential equation of first order 
\begin{equation}
\frac{\partial\hat p}{\partial\tau}+\sigma\bigl[1+D(\tau)\sigma\bigr]\frac{\partial\hat p}{\partial\sigma}=-\sigma m(\tau)\hat p,
\label{fpe_2}
\end{equation}
with initial condition [cf. Eq. (\ref{initial})]
\begin{equation}
\hat p(\sigma, 0|x_0)=e^{-\sigma x_0}.
\label{initial_2}
\end{equation}

The exact solution to this initial-value problem can be obtained by the method of characteristics \cite{courant}. This is detailed in Appendix \ref{AppA} with the result 
\begin{equation}
\hat p(\sigma,\tau|x_0)=\exp\Biggl\{-\frac{\sigma x_0e^{-\tau}}{1+\sigma\phi(\tau)}-
\sigma\int_0^{\phi(\tau)}\frac{\theta_\tau(\xi)d\xi}{1+\sigma\xi}\Biggr\},
\label{solution_2}
\end{equation}
where
\begin{equation}
\theta_\tau(\xi)\equiv\frac{m[\tau-\bar\tau(\xi)]}{D[\tau-\bar\tau(\xi)]},
\label{theta_tau}
\end{equation}
$\bar\tau(\xi)$ is implicitly defined by  
\begin{equation}
\xi=\int_0^{\bar\tau(\xi)} e^{-s} D(\tau-s) ds,
\label{tau(xi)}
\end{equation}
and the function $\phi(\tau)$ by
\begin{equation}
\phi(\tau)\equiv\int_0^{\tau} e^{-s}D(\tau-s)ds.
\label{phi_2}
\end{equation}

We next write the probability distribution in terms of the original time scale $t$. Note that obtaining the CF
$\hat p(\sigma,t|x_0,t_0)$ certainly amounts to substituting in Eq. (\ref{solution_2}) the dimensionless time scale $\tau$ by the function $\tau(t)$ given in Eq. (\ref{tau}). However, we also have to write $\theta_\tau(\xi)$ [cf. Eq. (\ref{theta_tau})] as a function of $t$ which is rather involved. In the Appendix \ref{real_time} we show that 
\begin{widetext}
\begin{equation}
\hat p(\sigma,t|x_0,t_0)=\exp\Biggl\{-\frac{\sigma x_0 e^{-\int_{t_0}^t \alpha(s)ds}}{1+\sigma\phi(t,t_0)} -
\sigma\int_0^{\phi(t,t_0)}\frac{\theta_t(\xi)}{1+\sigma\xi}d\xi\Biggr\},
\label{real_time_1}
\end{equation}
\end{widetext}
where,
\begin{equation}
\phi(t,t_0)\equiv\frac 12 \int_{t_0}^t k^2(t')\exp\left\{-\int_{t'}^t \alpha(s)ds\right\}dt',
\label{phi_t}
\end{equation}
\begin{equation}
\theta_t(\xi)\equiv\frac{2\beta[t(\xi)]}{k^2[t(\xi)]},
\label{theta(xi)}
\end{equation}
and $t(\xi)$ is defined by $\phi[t|t(\xi)]=\xi$. That is, 
\begin{equation}
\frac 12 \int_{t(\xi)}^t k^2(t')\exp\left\{-\int_{t'}^t \alpha(s)ds\right\}dt'=\xi.
\label{t(xi)}
\end{equation}

Equation (\ref{real_time_1}) is our main result and completely determines, through the characteristic function, the probability distribution of the non-stationary Feller process. However, the exact analytical inversion of this equation, thus obtaining the PDF, seems to be beyond reach except for few special cases to be detailed in the next section. It is, nonetheless, possible to get various approximate forms (see Secs. \ref{assym} and \ref{origin}) as well as obtaining exact expression for the cumulants of any order. 

In effect, in terms of the CF, cumulants $\kappa_n(t|t_0)$ are defined by
\begin{equation}
\kappa_n(t|t_0)=(-1)^n\left.\frac{\partial^n}{\partial\sigma^n}\ln\hat p(\sigma,t|x_0,t_0)\right|_{\sigma=0},
\label{cumulant_def}
\end{equation}
($n=1,2,3,\dots)$. Substituting for Eq. (\ref{real_time_1}), we write
$$
\kappa_n(t|t_0)=(-1)^{n+1}\left.\frac{\partial^n \hat f(\sigma, t|t_0)}{\partial\sigma^n}\right|_{\sigma=0},
$$
where
$$
\hat f(\sigma,t|t_0)=\frac{\sigma x_0e^{-\int_{t_0}^t \alpha(s)ds}}{1+\sigma\phi(t,t_0)}+
\sigma\int_0^{\phi(t,t_0)}\frac{\theta_t(\xi)}{1+\sigma\xi}d\xi.
$$
The expansion in powers of $\sigma$ yields
\begin{eqnarray*}
\hat f(\sigma,t|t_0)&=&\sum_{n=0}^\infty (-1)^n\biggl[x_0 \phi^n(t,t_0) e^{-\int_{t:_0}^t \alpha(s)ds}\\
&+&
\int_0^{\phi(t,t_0)} \xi^n \theta_t(\xi) d\xi\biggr]\sigma^{n+1},
\end{eqnarray*}
and we finally have
\begin{eqnarray}
\kappa_n(t|t_0)&=&n!\Biggl[x_0\phi^{n-1}(t,t_0) e^{-\int_{t_0}^t\alpha(s)ds}\nonumber \\ 
&+&\int_0^{\phi(t,t_0)} \xi^{n-1} \theta_t(\xi) d\xi\biggr],
\label{cumulant_1}
\end{eqnarray}
($n=1,2,3,\dots)$, as the exact expression for cumulants of any order.

\section{Some special cases}
\label{special}

We will now obtain exact expressions of the probability density function (PDF) in three particular instances. 
\bigskip

(i) Suppose first that  the drift of the processes also vanishes at the origin. In other words, $\beta(t)=0$ which is equivalent to $m(\tau)=0$ [cf. Eq. (\ref{rescaled_par})]. This is a rather peculiar situation because the normal level coincides with  the singular boundary meaning that the process tends to that singularity as time progresses (see discussion in Sec. \ref{dynamics}). This is the case treated in Ref. \cite{gan}. 

Now the parameter $\theta_\tau(\xi)$ defined in Eq. (\ref{theta(xi)}) also vanishes and the CF (\ref{real_time_1}) reads
\begin{equation}
\hat p(\sigma,t|x_0,t_0)=\exp\left\{-\frac{\sigma x_0 e^{-\tau(t)}}{1+\sigma\phi(t,t_0)}\right\},
\label{cf_0}
\end{equation}
where $\phi(t,t_0)$ is given by Eq. (\ref{phi_t}) and 
\begin{equation}
\tau(t)=\int_{t_0}^t \alpha(s) ds.
\label{tau(t)}
\end{equation}

In the Appendix \ref{AppB0} we show that the Laplace inversion of Eq. (\ref{cf_0}) is
\begin{widetext}
\begin{equation}
p(x,t|x_0,t_0)=e^{-x_0 e^{-\tau(t)}/\phi(t,t_0)}\delta(x)
+ \frac{e^{-\tau(t)/2}}{\phi(t,t_0)} \left(\frac{x_0}{x}\right)^{1/2} 
\exp\left\{-\frac{x+x_0e^{-\tau(t)}}{\phi(t,t_0)}\right\} I_1\left[2\frac{\sqrt{xx_0e^{-\tau(t)}}}{\phi(t,t_0)}\right],
\label{pdf_0}
\end{equation}
\end{widetext}
where $I_1(z)$ is a modified Bessel function.

Equation (\ref{pdf_0})  agrees with the PDF obtained by Gan and Waxman who have stressed the singular character of the origin \cite{gan} something absent in Feller's original formulation \cite{feller1}. Indeed, the term involving the Dirac function reflects the singularity of the origin. For, as we readily see from Eq. (\ref{pdf_0})  
$$
p(x,t|0,t_0)=\delta(x),
$$
which implies that if the process starts at $x_0=0$ remains there forever.

Cumulants are given by Eq. (\ref{cumulant_1}) which, since $\theta_{t}(\xi)=0$, read:
\begin{equation}
\kappa_n(t|t_0)=n! x_0 \phi^{n-1}(t,t_0) e^{-\tau(t)},
\label{cumulant_0}
\end{equation}
($n=1,2,3,\dots)$.

\bigskip

(ii) Suppose now that $\beta(t)$ and $k^2(t)$, are proportional to each other for all 
$t\geq t_0$. In this case the ratio $\theta_t(\xi)$ defined in Eq. (\ref{theta(xi)}) is constant:
\begin{equation}
\frac{2\beta(t)}{k^2(t)}=\theta,
\label{theta}
\end{equation}
where $\theta>0$ is a positive constant. Note incidentally that in the dimensionless time scale $\tau$ this case corresponds to having the normal level $m(\tau)$ proportional to the diffusion coefficient $D(\tau)$ for all $\tau\geq 0$.  

The characteristic function, Eq. (\ref{real_time_1}), now reads
$$
\hat p(\sigma,t|x_0,t_0)=\exp\Biggl\{-\frac{\sigma x_0e^{-\tau(t)}}{1+\sigma\phi(t,t_0)}-
\theta\sigma\int_0^{\phi(t,t_0)}\frac{d\xi}{1+\sigma\xi}\Biggr\}.
$$
The integral in the exponential is easily evaluated and we have
\begin{equation}
\hat p(\sigma,t|x_0,t_0)=
\frac{1}{\bigl[1+\sigma\phi(t,t_0)\bigr]^{\theta}}\exp\Biggl\{-\frac{\sigma x_0e^{-\tau(t)}}{1+\sigma\phi(t,t_0)}\Biggr\}.
\label{cf_constant}
\end{equation}

The Laplace inversion of this expression is sketched in the Appendix \ref{AppB} and the exact PDF is
\begin{widetext}
\begin{equation}
p(x,t|t_0,x_0) = \frac{1}{\phi(t,t_0)} \left(\frac{x}{x_0e^{-\tau(t)}}\right)^{(\theta-1)/2} 
\exp\left\{-\frac{x+x_0e^{-\tau(t)}}{\phi(t,t_0)}\right\} 
I_{\theta-1}\left[2\frac{\sqrt{xx_0e^{-\tau(t)}}}{\phi(t,t_0)}\right],
\label{pdf_cons}
\end{equation}
\end{widetext}
where $I_{\theta-1}(z)$ is a modified Bessel function of order $\theta-1$.

Cumulants are 
\begin{equation}
\kappa_n(t|t_0)=n!\left[x_0 \phi^{n-1}(t,t_0) e^{-\tau(t)}+\frac 1n \theta \phi^n(t,t_0)\right],
\label{cumulant_cons}
\end{equation}
($n=1,2,3,\dots)$.

\bigskip

(iii) Our last special case for which the PDF can be obtained exactly is when all parameters are time independent:
$$
\alpha(t)=\alpha, \quad \beta(t)=\beta, \quad k(t)=k,
$$
which corresponds to the original Feller process \cite{feller1}. Now 
\begin{equation}
\tau=\alpha(t-t_0), \quad \theta=\frac{2\beta}{k^2}, \quad D=\frac{k^2}{2\alpha},
\label{cons_para}
\end{equation}
and since $\theta$ is constant this case is a particular case of (ii) above. Therefore, the PDF of the stationary process will be given by Eq. (\ref{pdf_cons}) with 
[cf. Eq. (\ref{phi_2})]
$$
\phi(t,t_0)=D\left[1-e^{-(t-t_0)}\right].
$$
Note that since the process is now stationary we can set $t_0=0$ without loss of generality.

\section{Long-time asymptotics}
\label{assym}

In this section we get some interesting approximations to the PDF which are effective for sufficiently long times.  

Let us first remember that one of our basic assumptions, as expressed in Sec. \ref{dynamics}, is that the time-dependent parameter $\alpha(t)>0$ is such that if $t$ is long so is the new time scale $\tau$ defined in Eq. (\ref{tau}). In other words, 
$t\to\infty$ implies $\tau\to\infty$. 

We thus start off with the CF in the form given by Eq. (\ref{solution_2}):
$$
\hat p(\sigma,\tau|x_0)=\exp\Biggl\{-\frac{\sigma x_0e^{-\tau}}{1+\sigma\phi(\tau)}-
\sigma\int_0^{\phi(\tau)}\frac{\theta_\tau(\xi)d\xi}{1+\sigma\xi}\Biggr\}.
$$
Since $\tau$ is large,  the definition of $\theta_\tau(\xi)$ given in Eq. (\ref{theta_tau}) allows for the following 
approximation
\begin{equation}
\theta_\tau(\xi)\equiv\frac{m[\tau-\tau(\xi)]}{D[\tau-\tau(\xi)]}\simeq \frac{m(\tau)}{D(\tau)},
\label{theta_approx}
\end{equation}
$(\tau\gg 1)$, which enables us to estimate the integral appearing in the CF as
$$
\sigma\int_0^{\phi(\tau)}\frac{\theta_\tau(\xi)d\xi}{1+\sigma\xi}\simeq\theta(\tau)\ln\bigl[1+\sigma\phi(\tau)\bigr],
$$
where
\begin{equation}
\theta(\tau)\equiv\frac{m(\tau)}{D(\tau)}.
\label{theta(tau)}
\end{equation}
Hence, the  CF is approximated by
\begin{equation}
\hat p(\sigma, \tau|x_0)\simeq\frac{1}{\bigl[1+\sigma\phi(\tau)\bigr]^{\theta(\tau)}}
\exp\left\{-\frac{\sigma x_0 e^{-\tau}}{1+\sigma\phi(\tau)}\right\},
\label{cf_app1}
\end{equation}
$(\tau\gg 1)$. Note that this approximate CF has de same form than that of Eq. (\ref{cf_constant}). Therefore, the approximate PDF would be given by Eq. (\ref{pdf_cons}) with the constant parameter $\theta$ replaced by the time-varying $\theta(\tau)$. 

We also observe that, within the same degree of approximation for which Eq. (\ref{theta_approx}) holds, we may write [cf. Eq. (\ref{phi_2})]
\begin{eqnarray*}
\phi(\tau)&\equiv&\int_0^\tau e^{-s}D(\tau-s)ds\\
&\simeq& D(\tau)\int_0^\tau e^{-s}ds= D(\tau)\left(1-e^{-\tau}\right), 
\end{eqnarray*}
which, after neglecting the exponentially small term $e^{-\tau}$, yields
\begin{equation}
\phi(\tau)\simeq D(\tau), \qquad (\tau\gg 1).
\label{phi_approx}
\end{equation} 

If, as we did in obtaining Eq. (\ref{phi_approx}), we also neglect exponentially small terms in Eq. (\ref{cf_app1}), we get
\begin{equation}
\hat p(\sigma,\tau)\simeq\frac{1}{\bigl[1+\sigma D(\tau)\bigr]^{\theta(\tau)}},  
\label{cf_app2} 
\end{equation}
$(\tau\gg 1)$, where, since time is large, the dependence on the initial value $x_0$ has vanished.

On the other hand, 
$$
\frac{1}{\bigl[1+\sigma D(\tau)\bigr]^{\theta(\tau)}}=
\frac{1/[D(\tau)]^{\theta(\tau)}}{\bigl[\sigma +1/D(\tau)\bigr]^{\theta(\tau)}},
$$
and using the Laplace inversion formula \cite{roberts}
$$
\mathcal{L}^{-1}\left\{\frac{1}{(\sigma+a)^{\gamma}}\right\}=\frac{1}{\Gamma(\gamma)}x^{\gamma-1} e^{-ax},
$$
$(\gamma>0)$ we obtain for $\tau\gg 1$ the following approximation
\begin{equation}
p(x,\tau)\simeq\frac{1}{\Gamma[\theta(\tau)]}\frac{x^{\theta(\tau)-1}}{[D(\tau)]^{\theta(\tau)}} e^{-x/D(\tau)}.
\label{pdf_asym}
\end{equation}

The PDF in real time, $p(x,t)$,  will be given by Eq. (\ref{pdf_asym}) after replacing $\tau$ by the function $\tau(t)$ given in 
Eq. (\ref{tau(t)}). Moreover, from Eqs. (\ref{rescaled_par}) and (\ref{theta(tau)}) we obtain 
\begin{equation}
\theta[\tau(t)]=\frac{m[\tau(t)]}{D[\tau(t)]}=\frac{2\beta(t)}{k^2(t)}\equiv\theta(t),
\label{theta(t)}
\end{equation}
and
\begin{equation}
D[\tau(t)]=\frac{k^2(t)}{2\alpha(t)}\equiv D(t).
\label{D(t)}
\end{equation}
Therefore
\begin{equation}
p(x,t)\simeq\frac{1}{\Gamma[\theta(t)]}\frac{x^{\theta(t)-1}}{[D(t)]^{\theta(t)}} e^{-x/D(t)},
\label{pdf_asym_t}
\end{equation}
$(t\gg t_0)$. We thus see that for long times the PDF is approximated by a Gamma distribution with shape and scale parameters, $\theta(t)$ and $D(t)$ respectively, dependent of time. 

Let us remark that the  asymptotic approximation given either by Eq. (\ref{pdf_asym}) or Eq. (\ref{pdf_asym_t}) does not apply to the singular case  $\beta(t)=0$ [which implies $\theta(t)=0$]. Indeed, in such a case Eq. (\ref{cf_app2}) reads $\hat p(\sigma,t)\simeq 1$ which after Laplace inversion yields
$$
p(x,t)\simeq \delta(x), \qquad (t\to\infty).
$$
Reflecting the fact that the homogeneous drift, $-\alpha(t) x$, drags the process towards the origin as $t\to\infty$. Let us incidentally  note that the same result is indeed achieved from the exact solution  (\ref{pdf_0}) in the limit $t\to\infty$. 

We end this section by getting the asymptotic expression of cumulants and moments. At first sight the most direct way of obtaining the long-time form of cumulants is starting from the general expression given in Eq. (\ref{cumulant_1}) and taking there the limit $t\to\infty$. However, it turns out to be much simpler to start off from the asymptotic expression of 
the CF given by Eq. (\ref{cf_app2}) and apply the definition (\ref{cumulant_def}). The asymptotic cumulants thus read
\begin{equation}
\kappa_n(t)\simeq (n-1)! \theta(t) D^n(t), \qquad (t\to\infty),
\label{cumulants_asym}
\end{equation}
$(n=1,2,3,\dots)$. The connection between cumulants and moments becomes very involved as $n$ increases \cite{gardiner}. It is again simpler the direct calculation of moments based on Eq. (\ref{cf_app2}). Indeed, since 
$$
\left\langle X^n(t)\right\rangle=\left. (-1)^n \frac{\partial^n \hat p}{\partial\sigma^n}\right|_{\sigma=0},
$$
$($n=1,2,3,\dots$)$, we obtain
\begin{equation}
\left\langle X^n(t)\right\rangle\simeq D^n(t)\theta(t)[\theta(t)+1]\cdots[\theta(t)+n-1],
\label{moments_asym}
\end{equation}
($t\to\infty$). Let us first suppose that $\theta(t)\gg 1$ as $t\to\infty$, then  
$\left\langle X^n(t)\right\rangle\simeq [D(t)\theta(t)]^n=\left[\beta(t)/\alpha(t)\right]^n$, that is [cf. Eq. (\ref{rescaled_par})],
\begin{equation}
\left\langle X^n(t)\right\rangle\simeq m^n(t),
\label{moment_1}
\end{equation}
and long-time moments are only determined by the normal level of the process. In the opposite case when  
$\theta(t)\ll 1$, long-time moments are determined by the diffusion coefficient:
\begin{equation}
\left\langle X^n(t)\right\rangle\simeq D^n(t)\theta(t).
\label{moment_2}
\end{equation}
Note that when $\theta(t)\to\theta$ tends to a finite constant [note that case (ii) of previous the section is a particular instance] also leads to moments dominated by the diffusion coefficient:
\begin{equation}
\left\langle X^n(t)\right\rangle\simeq \frac{\Gamma(n+\theta)}{\Gamma(\theta)}D^n(t). 
\label{moment_3}
\end{equation}

\section{Near  the origin}
\label{origin}

Another situation where it is possible to know an approximate PDF is when the process is near the origin.  The behavior of the process for small values of $x$ is rather significant in many fields such as, for instance, the firing of neurons \cite{lanska} and also in population dynamics  where, as we mentioned before, attaining $x=0$ means extinction \cite{capocelli_ricciardi,capocelli,azaele}. 

The problem is closely related to the question of whether or not the non-stationary Feller process  can access the singular boundary $x=0$. For the stationary process this problem was addressed by Feller himself who, after analyzing the PDF near $x=0$, concluded that  if the constant ratio $\theta =2\beta/k^2\leq 1$ the origin is an accessible boundary, while if 
$\theta>1$ it is not \cite{feller1} (see also Ref. \cite{maso12}). In the stationary process the accessible character of $x=0$ is time independent and fixed forever once we know the parameters of the model.  We will now show that for the non-stationary process attaining the origin is not static but time dependent. 

Let us incidentally note that the question of reaching the origin would be more elegantly addressed by evaluating, instead of the PDF, the hitting probability of first reaching $x=0$, as we recently did for the stationary process \cite{maso12}. However, obtaining hitting probabilities needs the knowledge of first-passage time densities, something rather involved for non-stationary processes. We will, therefore, analyze the behavior of the PDF near the origin as was done by Feller \cite{feller1} . 

The starting point of our asymptotic analysis is the characteristic function in the form given by Eq.  (\ref{real_time_1})  and we will use this exact expression for knowing the behavior of the PDF near the origin. The later inference is achieved after using Tauberian theorems which show that the small $x$ behavior of $p(x,t|x_0,t_0)$ is determined by the large $\sigma$ behavior of its Laplace transform \cite{tauberian}. 

Following this procedure  we prove in  the Appendix \ref{AppC} that for large values of $\sigma$ we have 
\begin{equation}
\hat p(\sigma,t|x_0,t_0)=\frac{A(t|x_0,t_0)}{\sigma^{\theta(t)}}\left[1+O\left(\frac 1\sigma \ln\sigma\right)\right],
\label{asym_cf_1}
\end{equation}
where $A(t|x_0,t_0)$ is defined in Eq. (\ref{A}) of Appendix \ref{AppC} and  $\theta(t)$ in Eq. (\ref{theta(t)}). The CF is thus approximated by
$$
\hat p(\sigma,t|x_0,t_0)\simeq\frac{A(t|x_0,t_0)}{\sigma^{\theta(t)}}
$$
as $\sigma\to\infty$. Using Tauberian theorems \cite{tauberian} we see that the behavior of the PDF near the origin will be given by the Laplace inversion of this expression: 
\begin{equation}
p(x,t|x_0,t_0)\simeq K(t|x_0,t_0) x^{\theta(t)-1}, 
\label{power_law}
\end{equation}
$(x\to 0$ and $\theta(t)>0$), where 
$$
K(t|x_0,t_0)=\frac{A(t|x_0,t_0)}{\Gamma[\theta(t)]}.
$$

We, therefore, see that that the PDF near the origin follows a power law with a time-varying exponent and that at $x=0$ the PDF is different from zero only if $\theta(t)\leq 1$: 
\begin{equation}
p(0,t|x_0)= \begin{cases} \infty & \quad \theta(t)<1 \\ 
K(\tau|x_0) & \quad \theta(t)=1 \\ 
0 & \quad \theta(t)>1.
\end{cases}
\label{boundary1}
\end{equation}
Following Feller \cite{feller1} (see also \cite{maso12}) we conclude that if  $\theta(t)\leq 1$ the origin is an accessible boundary while if $\theta(t)>1$ it is not. Therefore, attaining the origin is not fixed but varies with time.

\section{Concluding remarks}
\label{conclusions}

We briefly summarize the main findings of the paper. In this work we have concentrated on the study of the non stationary Feller process with time-varying coefficients:
$$
dX(t)=-[\alpha(t) X(t)-\beta(t)]dt+k(t)dW(t),
$$
where $W(t)$ is the standard Wiener process. We have assumed: (i) $\alpha(t)>0$, $\beta(t)\geq 0$, and $k(t)>0$ are positive and smooth functions of $t\geq t_0$ where $t_0$ is an arbitrary initial time, and (ii) $\alpha(t)$ is such that
$$
\lim_{t\to\infty}\int_{t_0}^t \alpha(t')dt'=\infty.
$$

Under these rather general conditions on the coefficients, which ensure a broad applicability of the process, we have been able to obtain the exact probability distribution exemplified by the characteristic function of the process, as well as cumulants of any order [cf. Eqs. (\ref{real_time_1}) and (\ref{cumulant_1})]. 

We have been able to exactly invert the CF to get the PDF in a few special cases and obtain some relevant approximations as well. Thus, for sufficiently long times  the PDF approaches the Gamma distribution,
$$
p(x,t)\simeq\frac{1}{\Gamma[\theta(t)]}\frac{x^{\theta(t)-1}}{[D(t)]^{\theta(t)}}e^{-x/D(t)}. \qquad (t\gg t_0),
$$
with the shape of the distribution determined by 
$$
\theta(t)=\frac{2\beta(t)}{k^2(t)}>0,
$$ 
and the scale by 
$$
D(t)=\frac{k^2(t)}{2\alpha(t)},
$$
both parameters depending of time.

Another situation where it has been possible to get an approximate PDF is when the process is not far from the origin. Thus, for small values of the state variable $x$, the PDF obeys a power-law with a time-varying exponent,
$$
p(x,t|x_0,t_0)\sim x^{\theta(t)-1}, \qquad (x\to 0),
$$
where $\theta(t)>0$ is the ratio between the time-varying deterministic normal level $\beta(t)$ and the strength of fluctuations $k^2(t)/2$. Finally, the behavior near $x=0$ also determines the question, rather significant in some settings, of whether the origin is accessible or not . We have proved that for strong fluctuations such that $k^2(t)/2\geq \beta(t)$ the origin is accessible while for milder fluctuations  [{\it i.e.,} $k^2(t)/2<\beta(t)$] it is nor. The special character of the origin is now dynamical rather than static (as was the situation of the stationary case). 

We finish this work by mentioning few possible physical applications of the non-stationary process. One field is econophysics where the Feller process is one of the most popular models for the volatility \cite{heston,yakovenko}. As a first approximation the normal level of the volatility --{\it i.e.,} the parameter $m(t)\equiv\beta(t)/\alpha(t)$ [cf. Eq. (\ref{rescaled_par})]-- is assumed to be constant. However, there are evidences  of seasonality in the behavior of volatility \cite{wiggins,campbell} which could be modeled assuming a periodic function of time as normal level. A similar situation appears in the price of commodities  \cite{bollerslev,matia}. 

We can find other potential applications in anomalous diffusion problems. In the broad and complex field of fractional dynamics \cite{fractional_metzler,fractional_eliazar}, the non-stationary Feller process may represent a complementary  and rather  simple approximation to the problem that is not based on the fractional Brownian motion as driving noise. Thus, for instance, we have shown in Sec. \ref{assym} that when $\theta(t)\gg 1$ ($t\to\infty$) the asymptotic mean-square value of the process is $\langle X^2(t)\rangle\simeq m^2(t)$ [cf. Eq. (\ref{moment_1})]. If we further assume that as $t\to\infty$ the normal level follows a power law $m(t)\sim t^{\gamma/2}$ ($\gamma>0$) then the process shows the anomalous diffusion behavior: 
$$
\langle X^2(t)\rangle \sim t^\gamma. 
$$
It is quite interesting that, under the assumption $\theta(t)\gg 1$, the anomalous behavior depends only on the normal level $m(t)=\beta(t)/\alpha(t)$ and it is independent of the noise intensity $k^2(t)$. Other different situations may indeed appear depending on the log-time behavior of the parameters $\alpha(t), \beta(t)$ and $k(t)$ as we can see from the discussion at the end of Sec. \ref{assym} [cf. Eqs. (\ref{moment_2}) and (\ref{moment_3})].

In future works we will try to explore some of these possible applications.

\acknowledgments 

The author acknowledges partial financial support from MINECO under contracts no. FIS2013-47532-C3-2-P and FIS 2009-09689  and also from Generalitat de Catalunya under contract 2014 SGR 608.

\appendix

\section{The method of characteristics}
\label{AppA}

In this appendix we solve the initial-value problem (\ref{fpe_2})--(\ref{initial_2}) by the method of characteristics \cite{courant}.

We look for solutions of Eq. (\ref{fpe_2}) in the form
\begin{equation}
\hat p(\sigma,\tau|x_0)=e^{-\hat q(\sigma,\tau|x_0)}.
\label{q}
\end{equation}
If, in addition, we define the new variable
\begin{equation}
\eta=\frac 1\sigma,
\label{eta}
\end{equation}
problem (\ref{fpe_2})--(\ref{initial_2}) takes the simpler form:
\begin{equation}
\frac{\partial\hat q}{\partial\tau}-\bigl[\eta+D(\tau)\bigr]\frac{\partial\hat q}{\partial\eta}=\frac{m(\tau)}{\eta},
\label{a3}
\end{equation}
\begin{equation}
\hat q(\sigma, 0|x_0)=\frac{x_0}{\eta}.
\label{a4}
\end{equation}

The equations of the characteristics associated to Eq. (\ref{a3}) are \cite{courant}
\begin{equation}
d\tau=-\frac{d\eta}{\eta+D(\tau)}=\frac{d\hat q}{m(\tau)/\eta},
\label{characteristics}
\end{equation}
from which we have
$$
\frac{d\eta}{d\tau}=\eta-D(\tau).
$$
Hence
\begin{equation}
\eta(\tau)=e^{-\tau}\bigl[C_1-\psi(\tau)\bigr],
\label{eta_tau}
\end{equation}
where $C_1$ is an arbitrary constant and
\begin{equation}
\psi(\tau)\equiv\int_0^\tau e^{\tau'}D(\tau')d\tau'.
\label{psi}
\end{equation}

On the other hand, from Eq. (\ref{characteristics}) we also have $d\hat q/d\tau=m(\tau)/\eta$ which, after using  the expression for $\eta$ given in Eq. (\ref{eta_tau}), yields
$$
\frac{d\hat q}{d\tau}=\frac{e^{\tau}m(\tau)}{C_1-\psi(\tau)}.
$$ 
Therefore,
\begin{equation}
\hat q=C_2+\int_0^\tau \frac{e^{\bar\tau}m(\bar\tau)}{C_1-\psi(\bar\tau)}d\bar\tau,
\label{q_app}
\end{equation}
where $C_2$ is another arbitrary constant. 

Following the method of characteristics, the general solution of Eq. (\ref{a3}) is given by 
$$ 
C_2=F(C_1),
$$
where $F$ is an arbitrary function \cite{courant}. Thus, since [cf. Eqs (\ref{eta_tau}) and (\ref{q_app})]
\begin{equation}
C_1=e^\tau\eta+\psi(\tau), \quad C_2=\hat q-\int_0^\tau \frac{e^{\bar\tau}m(\bar\tau)}{C_1-\psi(\bar\tau)}d\bar\tau,
\label{c_1}
\end{equation}
we have
\begin{equation}
\hat q(\tau)=\int_0^\tau \frac{e^{\bar\tau}m(\bar\tau)}{C_1-\psi(\bar\tau)}d\bar\tau+F\left[e^{\tau}\eta+\psi(\tau)\right].
\label{a9}
\end{equation}

From the initial condition (\ref{a4}) and taking into account that $\psi(0)=0$, the unknown function $F$ reads 
$$
F(\eta)=\frac{x_0}{\eta},
$$
which substituting back into Eq. (\ref{a9}) yields
\begin{equation}
\hat q(\eta,\tau|x_0)=\int_0^\tau \frac{e^{\bar\tau}m(\bar\tau)}{C_1-\psi(\bar\tau)}d\bar\tau+\frac{x_0}{e^{\tau}\eta+\psi(\tau)}.
\label{a11}
\end{equation}
Recall that $C_1=e^\tau\eta+\psi(\tau)$ [cf. Eq. (\ref{c_1})], hence
$$
\hat q(\eta,\tau|x_0)=\int_0^\tau \frac{e^{\bar\tau}m(\bar\tau)}{e^{\tau}\eta+\psi(\tau)-\psi(\bar\tau)}d\bar\tau
+\frac{x_0}{e^{\tau}\eta+\psi(\tau)},
$$
but [cf. Eq. (\ref{psi})]
\begin{equation}
\psi(\tau)-\psi(\bar\tau)=\int_{\bar\tau}^\tau e^{\tau'}D(\tau')d\tau'\equiv\psi(\tau,\bar\tau),
\label{psi_tau}
\end{equation}
and we write Eq. (\ref{a11}) in the form
\begin{eqnarray*}
\hat q(\eta,\tau|x_0)&=&\frac{e^{-\tau}x_0}{\eta+e^{-\tau}\psi(\tau)}+
\int_0^\tau \frac{e^{-(\tau-\bar\tau)}m(\bar\tau)}{\eta+e^{-\tau}\psi(\tau,\bar\tau)}d\bar\tau\\
&=&\frac{e^{-\tau}x_0}{\eta+e^{-\tau}\psi(\tau)}+
\int_0^\tau \frac{e^{-\bar\tau}m(\tau-\bar\tau)}{\eta+e^{-\tau}\psi(\tau,\tau-\bar\tau)}d\bar\tau.
\end{eqnarray*}
However [cf. Eq. (\ref{psi_tau})]
\begin{eqnarray*}
e^{-\tau}\psi(\tau,\tau-\bar\tau)&=&\int_{\tau-\bar\tau}^\tau e^{-(\tau-\tau')}D(\tau')d\tau'\\
&=&\int_{0}^{\bar\tau} e^{-s}D(\tau-s)ds.
\end{eqnarray*}
Likewise [cf. Eq. (\ref{psi})]
$$
e^{-\tau}\psi(\tau)=\int_{0}^{\tau} e^{-s}D(\tau-s)ds.
$$
Hence,
\begin{equation}
\hat q(\eta,\tau|x_0)=\frac{e^{-\tau}x_0}{\eta+\phi(\tau)}+
\int_0^\tau \frac{e^{-\bar\tau}m(\tau-\bar\tau)}{\eta+\phi_\tau(\bar\tau)}d\bar\tau,
\label{q_1}
\end{equation}
where 
\begin{equation}
\phi_\tau(\bar\tau)\equiv\int_0^{\bar\tau} e^{-s}D(\tau-s) ds,
\label{phi_1_app}
\end{equation}
and
\begin{equation}
\phi(\tau)\equiv\phi_\tau(\tau)=\int_0^{\bar\tau} e^{-s}D(\tau-s) ds,
\label{phi_2_app}
\end{equation}

Back to the original variable $\sigma=1/\eta$ [cf. Eq. (\ref{eta})], the characteristic function reads [cfs. Eq. (\ref{q}) and (\ref{q_1})] 
\begin{equation}
\hat p(\sigma,\tau|x_0)=\exp\Biggl\{-\frac{\sigma x_0e^{-\tau}}{1+\sigma \phi(\tau)}-
\sigma \int_0^\tau \frac{e^{-\bar\tau}m(\tau-\bar\tau)}{1+\sigma \phi_\tau(\bar\tau)}d\bar\tau\Biggr\}.
\label{solution}
\end{equation}

Another convenient representation of the solution is obtained as follows. In the integral appearing in the right hand side of Eq. (\ref{solution}) we define a new integration variable
$$
\xi=\phi_\tau(\bar\tau),
$$
so that [cf. Eq. (\ref{phi_1_app})] 
\begin{equation}
d\xi=e^{-\bar\tau}D(\tau-\bar\tau)d\bar\tau.
\label{dxi}
\end{equation}
Moreover, $\bar\tau=0$ implies $\xi=0$, while $\bar\tau=\tau$ implies $\xi=\phi(\tau)$. Hence,
\begin{eqnarray*}
&&\int_0^\tau \frac{e^{-\bar\tau}m(\tau-\bar\tau)}{1+\sigma\phi_\tau(\bar\tau)}d\bar\tau \\
&&=\int_0^{\phi(\tau)}\frac{m[\tau-\bar\tau(\xi)]}{D[\tau-\bar\tau(\xi)]} \frac{d\xi}{1+\sigma\xi}.
\end{eqnarray*}
where $\bar\tau(\xi)$ is implicitly defined defined  by $\xi=\phi_{\tau}[\bar\tau(\xi)]$, that is, 
\begin{equation}
\int_0^{\bar\tau(\xi)} e^{-s}D(\tau-s)ds=\xi.
\label{tau(xi)_app}
\end{equation}
Therefore, defining
\begin{equation}
\theta_\tau(\xi)\equiv\frac{m[\tau-\bar\tau(\xi)]}{D[\tau-\bar\tau(\xi)]},
\label{theta_tau_app}
\end{equation}
we prove Eq. (\ref{solution_2}):
\begin{equation}
\hat p(\sigma,\tau|x_0)=\exp\Biggl\{-\frac{\sigma x_0e^{-\tau}}{1+\sigma\phi(\tau)}-
\sigma\int_0^{\phi(\tau)}\frac{\theta_\tau(\xi)d\xi}{1+\sigma\xi}\Biggr\}.
\label{solution_2_app}
\end{equation}

\section{Distributions of probability in the original time scale}
\label{real_time}

We start off with Eq. (\ref{solution}) of Appendix \ref{AppA} which after a simple change of variables in the integral reads
\begin{eqnarray}
\hat p(\sigma,\tau|x_0)&=&\exp\Biggl\{-\frac{\sigma x_0e^{-\tau}}{1+\sigma\phi(\tau)}\nonumber \\
&-&\sigma\int_0^\tau\frac{m(\tau')e^{-(\tau-\tau')}}{1+\sigma\phi_\tau(\tau-\tau')}d\tau'\Biggr\},
\label{solution_app}
\end{eqnarray}
where $\phi_\tau(\tau-\tau')$ and $\phi(\tau)$ are given by  Eqs. (\ref{phi_1_app}) and (\ref{phi_2_app}) which we write in the form
\begin{equation}
\phi_\tau(\tau-\tau')=\int_{\tau'}^{\tau} e^{-(\tau-\tau'')}D(\tau'')d\tau'',
\label{phi_1_appB}
\end{equation}
and $\phi(\tau)\equiv\phi_\tau(\tau)$. 

Denote by $I(\tau)$ the integral appearing in Eq. (\ref{solution_app}):
$$
I(\tau)\equiv \int_0^\tau\frac{m(\tau')e^{-(\tau-\tau')}}{1+\sigma\phi_\tau(\tau-\tau')}d\tau'.
$$
 Bearing in mind Eq. (\ref{tau}) we define a new integration variable $t'$ by 
\begin{equation}
\tau'=\tau(t')=\int_{t_0}^{t'}\alpha(t'')dt''.
\label{tau'}
\end{equation}
Thus $d\tau'=\alpha(t')dt'$ and $\tau'=0$ implies $t'=t_0$ whereas $\tau'=\tau$ --{\it i.e.,} $\tau(t')=\tau(t)$-- implies $t'=t$. Hence
\begin{equation}
I(t)\equiv I[\tau(t)]=\int_{t_0}^t\frac{m[\tau(t')]\alpha(t') e^{-[\tau(t)-\tau(t')]}}{1+\sigma\phi_{\tau(t)}[\tau(t)-\tau(t')]}dt'.
\label{inter_1}
\end{equation}
However, from the definition of $m(\tau)$ given in Eq. (\ref{rescaled_par}) we see that $m[\tau(t')]=\beta(t')/\alpha(t')$ and
\begin{equation}
m[\tau(t')]\alpha(t')=\beta(t').
\label{inter_2}
\end{equation}

On the other hand, from Eq. (\ref{phi_1_app}) we have
$$
\phi_{\tau(t)}[\tau(t)-\tau(t')]=e^{-\tau(t)}\int_{\tau(t')}^{\tau(t)} e^{\tau''} D(\tau'')d\tau'', 
$$
thus performing again the change of variable $\tau''=\tau(t'')$, we get [cf. Eqs. (\ref{tau}) and (\ref{rescaled_par})]
\begin{eqnarray}
&&\phi_{\tau(t)}[\tau(t)-\tau(t')]=e^{-\tau(t)}\int_{t'}^{t} e^{\tau(t'')} D[\tau(t'')]\alpha(t'')dt'' \nonumber\\
&&=\frac 12 \int_{t'}^{t} k^2(t'') \exp\left\{-\int_{t''}^t \alpha(s)ds\right\} dt'' \nonumber\\
&\equiv& \phi(t|t')
\label{inter_3},
\end{eqnarray}
where we used Eq. (\ref{rescaled_par}) to write
\begin{equation}
D[\tau(t'')]\alpha(t'')=\frac 12 k^2(t'').
\label{inter_4}
\end{equation}
Note also that in the original time scale, $\phi(\tau)\equiv\phi_\tau(\tau)$ will be given by $\phi[\tau(t)]=\phi(t,t_0)$, where 
[cf. Eq. (\ref{inter_3})]
\begin{equation}
\phi(t,t_0)=\frac 12 \int_{t_0}^{t} k^2(t') \exp\left\{-\int_{t'}^t \alpha(s)ds\right\} dt'.
\label{phi_t_app}
\end{equation}

Substituting Eqs. (\ref{inter_2}) and (\ref{inter_3}) into Eq. (\ref{inter_1}) yields
$$
I(t)=\int_{t_0}^t\frac{\beta(t') e^{-[\tau(t)-\tau(t')]}}{1+\sigma\phi(t|t')}dt'.
$$

The last step consists in performing the change of integration variable $\xi=\phi(t|t')$, that is,
$$
\xi=\frac 12 \int_{t'}^t e^{-[\tau(t)-\tau(t'')]} k^2(t'')dt'',
$$
whence 
$$
dt'=-\frac{2}{k^2(t')} e^{[\tau(t)-\tau(t')]}d\xi.
$$
Moreover, $t'=t$ implies $\xi=0$ whereas $t'=t_0$ implies $\xi=\phi(t,t_0)$. Therefore,
$$
I(t)=\int_{0}^{\phi(t,t_0)}\frac{2\beta(t')}{k^2(t')}\frac{d\xi}{1+\sigma\xi},
$$
where $t'=t(\xi)$ is implicitly defined by [see Eqs. (\ref{tau}) and (\ref{tau'})]
\begin{equation}
\xi=\frac 12 \int_{t(\xi)}^t  k^2(t'') \exp\left\{-\int_{t''}^t \alpha(s)ds\right\}dt''.
\label{implicit_app}
\end{equation}

Finally, defining
\begin{equation}
\theta_t(\xi)\equiv\frac{2\beta[t(\xi)]}{k^2[t(\xi)]},
\label{theta_t_app}
\end{equation}
we write
\begin{equation}
I(t)=\int_{0}^{\phi(t,t_0)}\frac{\theta_t(\xi)}{1+\sigma\xi}d\xi,
\label{I(t)_app}
\end{equation}
and collecting results we get
\begin{eqnarray}
\hat p(\sigma,t|x_0,t_0)&=&\exp\Biggl\{-\frac{\sigma x_0e^{-\int_{t_0}^t \alpha(s)ds}}{1+\sigma\phi(t,t_0)}\nonumber \\
&-&\sigma\int_0^{\phi(t,t_0)}\frac{\theta_t(\xi)}{1+\sigma\xi}d\xi\Biggr\},
\label{solution_t_app}
\end{eqnarray}
which is Eq. (\ref{real_time_1}).

\section{Proof of Eq. (29)}
\label{AppB0}

We start off with Eq. (\ref{cf_0}):
\begin{equation}
\hat p(\sigma,t|x_0,t_0)=\exp\left\{-\frac{\sigma x_0 e^{-\tau}}{1+\sigma\phi}\right\},
\label{c01}
\end{equation}
where $r=r(t)$ and $\phi=\phi(t,t_0)$ are defined in Eqs. (\ref{tau(t)}) and (\ref{phi_t}) respectively. Since
$$
\frac{\sigma x_0 e^{-\tau}}{1+\sigma\phi}=\frac{x_0 e^{-\tau}}{\phi}-\frac{x_0 e^{-\tau}/\phi^2}{\sigma+1/\phi},
$$
the CF can be written as 
$$
\hat p(\sigma,t|x_0,t_0)=e^{-x_0 e^{-\tau}/\phi}\exp\left\{\frac{ x_0 e^{-\tau}/\phi^2}{\sigma+1/\phi}\right\},
$$
which after a power-law expansion yields
\begin{equation}
\hat p(\sigma,t|x_0,t_0)=e^{-x_0 e^{-\tau}/\phi}\left[1+\sum_{n=1}^\infty\frac{(x_0e^{-\tau}/\phi^2)^n}{n!(\sigma+1/\phi)^n}\right].
\label{c02}
\end{equation}

We denote by $\mathcal{L}^{-1}\{\cdot\}$ the inverse Laplace transform. Then, since \cite{roberts}
$$
\mathcal{L}^{-1}\{1\}=\delta(x),
$$
and
$$
\mathcal{L}^{-1}\left\{\frac{1}{[\sigma+1/\phi(t,t_0)]^n}\right\}=
\frac{1}{\Gamma(n)}x^{n-1}e^{-x/\phi(t,t_0)},
$$
the inversion of Eq. (\ref{c02}) reads
\begin{eqnarray}
 p(x,t|x_0,t_0)&=&e^{-x_0 e^{-\tau}/\phi}\Biggl[\delta(x) \nonumber\\ 
 &+&
 e^{-x/\phi}\sum_{n=1}^\infty \frac{(x_0e^{-\tau}/\phi^2)^n}{n! \Gamma(n)} x^{n-1}\Biggr].
\label{c03}
\end{eqnarray}
But (recall $x$ and $x_0$ are nonnegative) 
\begin{equation*}
\left(\frac{x_0 e^{-\tau}}{\phi^2}\right)^n x^{n-1}=\frac{1}{\phi}\sqrt{\frac{x_0 e^{-\tau}}{x}}
\left(\frac{\sqrt{xx_0e^{-\tau}}}{\phi}\right)^{2n-1},
\end{equation*}
and
\begin{eqnarray*}
&&\sum_{n=1}^\infty \frac{(x_0e^{-\tau}/\phi^2)^n }{n! \Gamma(n)} x^{n-1}\\
&&=\frac{1}{\phi}\sqrt{\frac{x_0 e^{-\tau}}{x}}\sum_{k=0}^\infty\frac{[\sqrt{xx_0e^{-\tau}}/\phi]^{2k+1}}
{(k+1)!\Gamma(k+1)}\\
&&=\frac{1}{\phi} \sqrt{\frac{x_0 e^{-\tau}}{x}} I_1\left(2\frac{\sqrt{xx_0e^{-\tau}}}{\phi}\right),
\end{eqnarray*}
where
$$
I_1(z)=\sum_{n=0}^\infty \frac{(z/2)^{2n+1}}{(n+1)!\Gamma(n+1)}
$$
is the modified Bessel function of first order \cite{mos}. Collecting results into Eq. (\ref{c03}) we get
\begin{widetext}
\begin{equation}
\hat p(\sigma,t|x_0,t_0)=e^{-x_0 e^{-\tau}/\phi}\delta(x)
+ \frac{e^{-\tau/2}}{\phi} \left(\frac{x_0}{x}\right)^{1/2} 
\exp\left\{-\frac{x+x_0e^{-\tau}}{\phi}\right\} I_1\left(2\frac{\sqrt{xx_0e^{-\tau}}}{\phi}\right), 
\label{pdf_0_app}
\end{equation}
\end{widetext}
which is Eq. (\ref{pdf_0}) of the main text.

\section{Proof of Eq. (33)}
\label{AppB}

When $\theta$ is constant the CF is given by Eq. (\ref{cf_constant}):
\begin{equation}
\hat p(\sigma,t|x_0,t_0)=
\frac{1}{\bigl(1+\sigma\phi \bigr)^{\theta}}\exp\Biggl\{-\frac{\sigma x_0e^{-\tau}}{1+\sigma\phi }\Biggr\},
\label{b0}
\end{equation}
where $\tau=\tau(t)$ and $\phi=\phi(t,t_0)$. Note that the exponential term equals the CF of the case discussed in Appendix \ref{AppB0}  [cf. Eq. (\ref{c01})]. We may, therefore, use Eq. (\ref{c02}) of Appendix \ref{AppB0} into Eq. (\ref{b0}) and write
\begin{equation*}
\hat p(\sigma,t|x_0,t_0)=\frac{e^{-x_0e^{-\tau}/\phi }}{(1+\sigma\phi)^\theta }
\left[1+\sum_{n=1}^\infty \frac{\bigl(x_0e^{-\tau}/\phi^2 \bigr)^n}{n! (\sigma+1/\phi )^{n}}\right].
\end{equation*}
That is,
\begin{equation}
\hat p(\sigma,t|x_0,t_0)=e^{-x_0e^{-\tau}/\phi }
\sum_{n=0}^\infty \frac{\bigl(x_0e^{-\tau}/\phi^2 \bigr)^n}{n! (\sigma+1/\phi )^{n+\theta}}.
\label{b1}
\end{equation}

Using \cite{roberts}
$$
\mathcal{L}^{-1}\left\{\frac{1}{(\sigma+1/\phi )^{n+\theta}}\right\}=
\frac{1}{\Gamma(\theta+n)}x^{n+\theta-1}e^{-x/\phi },
$$
($\theta>0$, $n=0,1,2,\dots$), the Laplace inversion of Eq. (\ref{b1}) reads
$$
p(x,\tau|x_0)=e^{-(x+x_0 e^{-\tau})/\phi }
\sum_{n=0}^\infty \frac{(x_0 e^{-\tau}/\phi^2 )^n}{n! \Gamma(\theta+n)} x^{n+\theta-1}.
$$
But (recall $x$ and $x_0$ are nonnegative) 
\begin{eqnarray*}
&&\left(\frac{x_0 e^{-\tau}}{\phi^2 }\right)^n x^{n+\theta-1}\\
&&=\phi^{\theta-1} \left(\sqrt{\frac{x}{x_0e^{-\tau}}}\right)^{\theta-1}
\left(\frac{\sqrt{xx_0e^{-\tau}}}{\phi }\right)^{2n+\theta-1},
\end{eqnarray*}
hence
\begin{eqnarray*}
p(x,t|x_0,t_0)&=&\frac{e^{-(x+x_0 e^{-\tau})/\phi }}{\phi }\left(\frac{x}{x_0e^{-\tau}}\right)^{(\theta-1)/2} \\
&\times& \sum_{n=0}^\infty \frac{(\sqrt{xx_0e^{-\tau}}/\phi)^{2n+\theta-1}}{n! \Gamma(\theta+n)}.
\end{eqnarray*}

We recognize the series appearing in this equation as the expression of a modified Bessel function. Indeed
$$
I_\nu(z)=\sum_{n=0}^\infty \frac{(z/2)^{2n+\nu}}{n!\Gamma(n+\nu+1)}
$$
is the modified Bessel function of order $\nu$ \cite{mos}. Therefore,
\begin{eqnarray*}
p(x,t|x_0,t_0) &=& \frac{1}{\phi } \left(\frac{x}{x_0e^{-\tau}}\right)^{(\theta-1)/2} \\
&\times& \exp\left\{-\frac{x+x_0e^{-\tau}}{\phi }\right\} 
 I_{\theta-1}\left(2\frac{\sqrt{xx_0e^{-\tau}}}{\phi}\right),
\end{eqnarray*}
which is Eq. (\ref{pdf_cons}) of the main text.

\section{Proof of Eq. (50)}
\label{AppC}

We start off with Eq. (\ref{real_time_1}) for the CF:
\begin{equation}
\hat p(\sigma,t|x_0,t_0)=\exp\Biggl\{-\frac{\sigma x_0e^{-\tau(t)}}{1+\sigma\phi(t,t_0)}-
\sigma\int_0^{\phi(t,t_0)}\frac{\theta_t(\xi)d\xi}{1+\sigma\xi}\Biggr\},
\label{c1}
\end{equation}
where $\phi(t,t_0)$ is defined in Eq. (\ref{phi_t}), and 
\begin{equation}
\theta_t(\xi)\equiv\frac{2\beta[t(\xi)]}{k^2[t(\xi)]}.
\label{c2}
\end{equation}
Denote by $I(\sigma,t)$ the integral term of Eq. (\ref{c1}), the change of integration variable $\xi=z/\sigma$ yields
$$
I(\sigma,t)\equiv\sigma\int_0^{\phi(t,t_0)}\frac{\theta_t(\xi)}{1+\sigma\xi}d\xi=
\int_0^{\sigma\phi(t,t_0)}\frac{\theta_t(z/\sigma)}{1+z}dz.
$$
Assuming $\sigma\to\infty$ we expand $\theta_t(z/\sigma)$ in powers of $1/\sigma$:
$$
\theta_t(z/\sigma)=\theta_t(0)+\frac 1\sigma z\theta'_t(0)+ O(1/\sigma^2).
$$
Substituting this expansion into $I(\sigma,t)$ and integrating we have
\begin{eqnarray}
&&I(\sigma,t)=\theta_t(0) \ln\bigl[1+\sigma\phi(t,t_0)\bigr] \label{I}  \\
&&+ \frac 1\sigma\theta'_t(0)\Bigl[\sigma\phi(t,t_0)-\ln\bigl[1+\sigma\phi(t,t_0)\bigr]\Bigr]+ O(1/\sigma^2) \nonumber.
\end{eqnarray}

Before proceeding further let us note from Eq. (\ref{c2}) that 
$$
\theta_t(0)=\frac{2\beta[t(0)]}{k^2[t(0)]},
$$
but $\xi=0$ implies $t(0)=t$ [cf. Eq. (\ref{t(xi)})]. Hence
\begin{equation}
\theta_t(0)=\frac{2\beta(t)}{k^2(t)}\equiv \theta(t),
\label{theta(t)_2}
\end{equation}
[cf Eq. (\ref{theta(t)})]. Also from Eq. (\ref{c2}) we have
\begin{eqnarray*}
\theta'_t(\xi)&\equiv&\frac{d}{d\xi}\theta_t(\xi)=2\frac{d}{d\xi}\left(\frac{\beta[t(\xi)]}{k^2[t(\xi)]}\right)\\
&=&2\frac{dt(\xi)}{d\xi}\frac{\dot\beta[t(\xi)] k^2[t(\xi)]-2k[t(\xi)]\dot k[t(\xi)] \beta[t(\xi)]}{k^4[t(\xi)]},
\end{eqnarray*}
where $\dot\beta$ and $\dot k$ denote derivatives. Differentiating both sides of  Eq. (\ref{t(xi)}) we get
$$
\frac{dt(\xi)}{d\xi}=-\frac{2}{k^2[t(\xi)]} \exp\left\{-\int_{t(\xi)}^t \alpha(s) ds\right\},
$$
and, since $t(0)=t$, we obtain
$$
\left.\frac{dt(\xi)}{d\xi}\right|_{\xi=0}=-\frac{2}{k^2(t)}.
$$
Therefore,
\begin{equation}
\theta'_t(0)=\frac{4}{k^5(t)}\left[2\dot k(t)\beta(t)-\dot\beta(t)k(t)\right].
\label{theta'}
\end{equation}
 
Let us resume the proof of Eq. (\ref{asym_cf_1}). Since $\ln\sigma/\sigma\to 0$ as $\sigma\to\infty$, the most important contributions to the value of $I(\sigma,t)$ in Eq. (\ref{I}) are
$$
I(\sigma,t)=\theta(t) \ln[1+\sigma\phi(t,t_0)] + \theta'_t(0)\phi(t,t_0)+O\left(\ln\sigma/\sigma\right).
$$
Plugging this into Eq. (\ref{c1}) we write 
\begin{eqnarray*}
\hat p(\sigma,t|x_0,t_0)&=&
\frac{e^{-\theta'_t(0)\phi(t,t_0)}}{[1+\sigma\phi(t,t_0)]^{\theta(t)}} \\
&\times&
\exp\left\{-\frac{\sigma x_0e^{-\tau(t)}}{1+\sigma\phi(t,t_0)}\right\}\left[1+O\left(\frac{\ln\sigma}{\sigma}\right)\right].
\end{eqnarray*}
But
$$
\frac{1}{[1+\sigma\phi(t,t_0)]^{\theta(t)}}=\frac{1}{\sigma^{\theta(t)}[\phi(t,t_0)]^{\theta(t)}}
\left[1+O\left(\frac 1\sigma\right)\right],
$$
and
$$
\exp\left\{-\frac{\sigma x_0e^{-\tau(t)}}{1+\sigma\phi(t,t_0)}\right\}=e^{-x_0e^{-\tau(t)}/\phi(t,t_0)}
\left[1+O\left(\frac 1\sigma\right)\right].
$$
Collecting results and bearing in mind that $1/\sigma\to 0$ faster than $\ln\sigma/\sigma$ as $\sigma\to\infty$, we finally obtain Eq. (\ref{asym_cf_1}):
\begin{equation}
\hat p(\sigma,t|x_0,t_0)=\frac{A(t|x_0,t_0)}{\sigma^{\theta(t)}}\left[1+O\left(\frac 1\sigma \ln\sigma\right)\right],
\label{cf_app}
\end{equation}
where
\begin{eqnarray}
&&A(t|x_0,t_0)=\frac{1}{[\phi(t,t_0)]^{\theta(t)}}\nonumber \\
&&\times \exp\Biggl\{-\theta'_t(0)\phi(t,t_0)
-\frac{x_0e^{-\tau(t)}}{\phi(t,t_0)}\Biggr\},
\label{A}
\end{eqnarray}
and $\theta'_t(0)$ given in Eq. (\ref{theta'}).

\end{document}